\documentclass[traditabstract]{aa} 
\usepackage{txfonts}
\usepackage{epstopdf}
\usepackage[switch]{lineno}
\usepackage{graphicx,longtable,lscape,natbib,amssymb,amssymb,amsmath}
\bibpunct{(}{)}{;}{a}{}{,} 
\newcommand{\ltsima} {$\; \buildrel < \over \sim \;$}  
\newcommand{\gtsima} {$\; \buildrel > \over \sim \;$}  
\newcommand{\lta} {\lower.5ex\hbox{\ltsima}}  
\newcommand{\gta} {\lower.5ex\hbox{\gtsima}}  
\newcommand{\Ha} {H$\alpha$}  
\newcommand{\Hb} {H$\beta$}

\newcommand{\ergshz}{\>{\rm erg}\,{\rm s}^{-1}\,{\rm Hz}^{-1}}

\newcommand{\kms}{$\rm{\,km \,s}^{-1}$}

\usepackage{color}

\begin{document}

\title{The MURALES survey. V.} \subtitle{Jet-induced star formation
  in 3C 277.3 (Coma A).  }

\author{Alessandro Capetti\inst{1} \and Barbara Balmaverde\inst{1} 
\and Clive Tadhunter\inst{2}\and Alessandro Marconi\inst{3,4} \and Giacomo Venturi\inst{14,4}
\and M. Chiaberge\inst{5,6} \and R.D. Baldi\inst{7} \and
S. Baum\inst{9} \and R. Gilli\inst{8} \and P. Grandi\inst{8} \and
Eileen T. Meyer\inst{12} \and G. Miley\inst{10} \and
C. O$'$Dea\inst{9} \and W. Sparks\inst{13} \and E. Torresi\inst{8}
\and G. Tremblay\inst{11}
} \institute {INAF - Osservatorio
  Astrofisico di Torino, Via Osservatorio 20, I-10025 Pino Torinese,
  Italy
  \and Department of Physics \& Astronomy, University of Sheffield, Sheffield S6 3TG, UK
\and Dipartimento di Fisica e Astronomia, Universit\`a di
Firenze, via G. Sansone 1, 50019 Sesto Fiorentino (Firenze), Italy
\and INAF - Osservatorio Astrofisico di Arcetri, Largo Enrico Fermi
5, I-50125 Firenze,Italy \and Space Telescope Science Institute,
3700 San Martin Dr., Baltimore, MD 21210, USA \and Johns Hopkins
University, 3400 N. Charles Street, Baltimore, MD 21218, USA \and
INAF- Istituto di Radioastronomia, Via Gobetti 101, I-40129 Bologna,
Italy \and INAF - Osservatorio di Astrofisica e Scienza dello Spazio
di Bologna, via Gobetti 93/3, 40129 Bologna, Italy \and Department
of Physics and Astronomy, University of Manitoba, Winnipeg, MB R3T
2N2, Canada \and Leiden Observatory, Leiden University, PO Box 9513,
NL-2300 RA, Leiden, the Netherlands \and Harvard-Smithsonian Center
for Astrophysics, 60 Garden St., Cambridge, MA 02138, USA \and
University of Maryland Baltimore County, 1000 Hilltop Circle,
Baltimore, MD 21250, USA
\and SETI Institute, 189 N. Bernado Ave Mountain View,CA 94043
\and Instituto de Astrof\'isica, Facultad de F\'isica, Pontificia Universidad Cat\'olica  de Chile, Casilla 306, Santiago 22, Chile
}

\date{} 

\abstract{We present observations obtained with the VLT/MUSE optical
  integral field spectrograph of the radio source 3C277.3, located at
  a redshift of 0.085 and associated with the galaxy Coma A. An
  emission line region fully enshrouds the double-lobed radio source, which is
  $\sim$60 kpc $\times$ 90 kpc in size. Based on the emission line
  ratios, we identified five compact knots in which the gas ionization
  is powered by young stars located as far as $\sim$ 60 kpc from the
  host. The emission line filaments surrounding the radio emission are
  compatible with ionization from fast shocks (with a velocity of
  350-500 \kms), but a contribution from star formation occurring at
  the edges of the radio source is likely. Coma A might be a unique
  example in the local Universe in which the expanding outflow
  triggers star formation throughout the whole radio source.}


\titlerunning{The MURALES survey. V: Jet-induced star formation in 3C
  277.3 (Coma A).  }  \authorrunning{A. Capetti et al.}  \maketitle

\section{Introduction}
\label{intro}

Active galactic nuclei (AGN) play a key role in the so-called feedback
process, that is, the exchange of matter and energy between AGN, their
host galaxies, and clusters of galaxies. In particular, relativistic
jets in radio-loud AGN interact violently with the external medium
(see, e.g., \citealt{hardcastle20} for a review). Evidence of this
interaction is often seen in local radio galaxies (RGs), in which
cavities are observed in the hot external gas filled by the
radio-emitting plasma (e.g., \citealt{birzan12}). However, we still
lack a comprehensive view of the effects that highly energetic jets
have on the host and its immediate environment; for example, it is not
understood how precisely the coupling between radio jets and ionized
gas occurs, and whether the jets can accelerate the gas above the host
escape velocity \citep{mcnamara07}.  In addition, it is still unclear
under which conditions jets enhance or quench star formation (positive
or negative feedback), which is an essential ingredient for
understanding the effects of the nuclear activity on the star
formation history and evolution of their host galaxies.

Positive feedback, that is, star formation triggered by jets or by the
expansion of radio lobes, has been reported in a few individual cases
at low redshifts: young stars have been detected in the filaments
along the jet of Centaurus~A
\citep{mould00,rejkuba02,crockett12,neff15,santoro15,salome16}, in an
object along the path of the jet in NGC~541 (the Minkowski object,
\citealt{croft06}), and at the termination of the jet of NGC~5643
\citep{cresci15}. Conversely, positive feedback appears to be
important in high-redshift radio galaxies (see
\citealt{miley08,odea21} for a review and, e.g.,
\citealt{steinbring14,steinbring11,gilli19} for recent results). The
analysis of the far-infrared spectral energy distribution of a
complete sample of z$>$1 3CR sources indicates that $\sim$40\% of them
are undergoing episodes of star formation with rates of hundreds of
solar masses per year \citep{podigachoski15}. The mean specific star
formation rate of RGs at z$>$2.5 is higher than in typical starforming
galaxies over the same redshift range \citep{drouart14}. This suggests
that positive feedback was more effective in earlier epochs, but it
might also just reflect the rich gaseous environments around such
sources if they are triggered by major mergers.

In this framework, the MUse RAdio-Loud Emission lines Snapshot
(MURALES) survey was carried out with the integral field spectrograph
MUSE at the Very Large Telescope (VLT) on a sample of 3CR radio
sources in order to explore the connection between ionized gas and the
relativistic jet plasma. We observed 37 radio galaxies with $z<0.3$
and $\delta <20^\circ$ and presented the main results in
\citet{balmaverde19} and \citet{balmaverde21}. Thanks to their
unprecedented depth (the median 3$\sigma$ surface brightness
  limit in the emission line maps is 6$\times$10$^{-18}$ erg s$^{-1}$
  cm$^{-2}$ arcsec$^{-2}$), these observations reveal emission line
regions extending by several dozen kiloparsec in most objects. We
explored the ionization mechanism of the ionized gas and its
connection (from the point of view of its distribution and
kinematics) with the radio jets.

The radio source 3C277.3 is associated with the galaxy Coma A. Despite
its name, Coma A is not associated with the Coma cluster, which is
located at a redshift of z=0.085 at the center of a group of galaxies
\citep{worrall16}.  Given its declination ($\delta = 27.6^\circ$), it
is formally not part of the sample selected for MURALES, but it was
included in the target list because previous observations
\citep{miley81,tadhunter00} revealed that a nebula of
ionized gas surrounds the radio source. This made 3C~277.3 an ideal
target for exploring the effect of the AGN on the surrounding medium. This is
the primary goal of MURALES. \citet{solorzano03} already obtained
integral field observations at two locations of Coma A, but with a
smaller field of view (14\farcs6 $\times$ 11\farcs3) than is covered by MUSE (1\arcmin $\times$ 1\arcmin). They reported a HII region east of the nucleus.

Coma A is an elliptical galaxy with a total absolute magnitude (measured
from the 2MASS images) of M$_K =-25.12$ with no signs of recent
interactions \citep{capetti00,madrid06}.  The central stellar velocity
dispersion derived from the Sloan Digital Sky Survey (SDSS) spectrum is $\sigma_s=196\pm$7 \kms,
corresponding to a mass of the supermassive black hole of M$_{\rm
  BH}\sim1.6 \times 10^8$ M$_\odot$ when the relation
between M$_{\rm BH}$ and $\sigma_s$ from \citet{gutelkin09} is adopted. The
active nucleus has optical emission line ratios typical of high-excitation galaxies (HEGs, \citealt{buttiglione10}).

The source 3C277.3 has an edge-brightened double-lobed Fanaroff-Riley
type II radio structure that extends over $\sim$ 90 kpc
\citep{vanbreugel85}, with a radio luminosity of P=1.6 10$^{33}
\ergshz$ at 178 MHz \citep{laing80}. \citet{knuettel19} studied its
polarization properties and concluded that the observed depolarization
is consistent with being produced by a foreground screen of ionized
gas. The authors derived a magnetic field strength of $\sim 1
\mu$G. The high depolarization in the northern lobe is likely due to
the Laing–Garrington effect \citep{garrington88}, which implies that
the northern jet is receding.

We adopt the following set of cosmological parameters:
$H_{\rm o}=69.7$ \kms\ Mpc$^{-1}$ and $\Omega_m$=0.286
\citep{bennett14}. At the redshift of 3C277.3, 1\arcsec\ corresponds
to 1.7 kpc.

\section{Observations and data reduction}
\label{observation}
Two observations, with an exposure time of 980 seconds each, were
obtained with the VLT/MUSE spectrograph in wide-field mode with
nominal wavelength range (4800-9300 \AA) on 18 January 2019, with a
seeing of $\sim$ 0\farcs5. The science field was empty enough to allow
sky subtraction without dedicated sky observations. We used the ESO
MUSE pipeline (version 1.6.2) to obtain a fully reduced and calibrated
data cube \citep{weilbacher20}. The absolute accuracy of the flux
  calibration is 4-7\%, depending on the emission line in question, and
  it varies for about 5\% across the field of view.

\begin{figure}
  \includegraphics[width=0.45\textwidth]{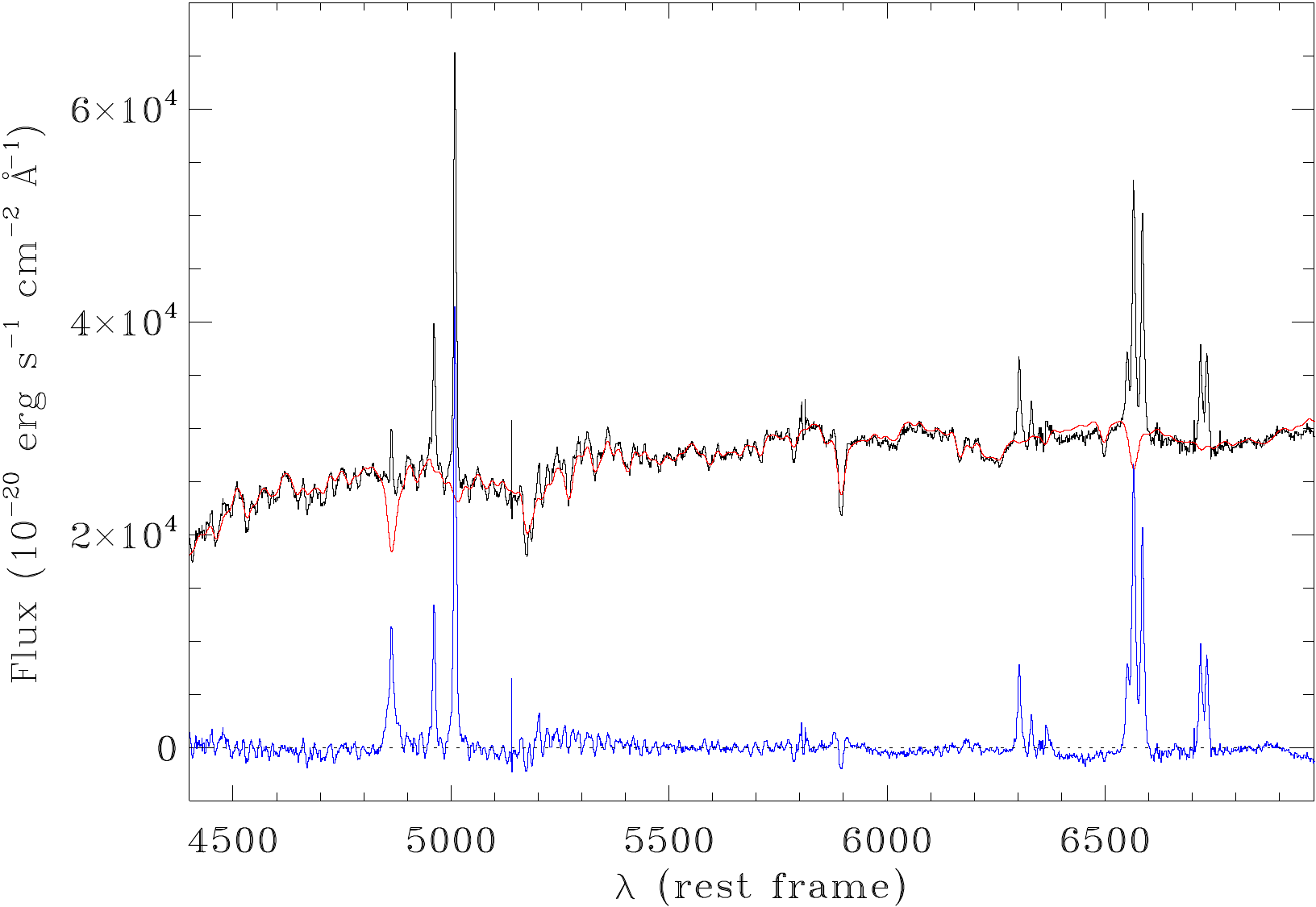}
  \caption{Example of subtraction of the stellar emission from the
    MUSE spectra. The black line is the spectrum extracted at the
    continuum peak, the red line is the best fit stellar continuum,
    the blue line represents the residuals after continuum subtraction
    showing the emission lines.}
\label{nuc}
\end{figure}

We followed the strategy for the data analysis described in
\citet{balmaverde19}. To summarize, we subtracted the stellar continuum using the MILES stellar templates library \citep{vazdekis10} after
resampling the data cube with Voronoi adaptive spatial binning
\citep{cappellari03}, requiring an average signal-to-noise ratio per
wavelength channel of at least 50 and using the penalized
pixel-fitting code (pPXF, \citealt{cappellari04}). As an example,
Fig. \ref{nuc} shows the result of the continuum subtraction on the
nucleus.

We then fitted the brightest emission lines (namely,
H$\beta$, [O~III]$\lambda\lambda$4959,5007,
[O~I]$\lambda\lambda$6300,6363, H$\alpha$,
[N~II]$\lambda\lambda$6548,6584, and [S~II]$\lambda\lambda$6716,6731)
in the continuum-subtracted spectra. We assumed that the lines in the
red and blue portion of the spectra have the same velocity
profile. While a single Gaussian reproduces the lines
profiles at large radii accurately, in the nuclear regions (i.e., within
2\arcsec\ from the galaxy center), we included an additional Gaussian
component. In the following figures, we only show the results
for the spaxels in which the line of interest is detected at a
2$\sigma$ level at least.

We re-reduced 1981 March L band, A-configuration VLA observations of
3C277.3 that were originally published in \citet{vanbreugel85}.  Data reduction
was carried out in CASA version 5.6.2-3, and the original uvfits files
were imported to a measurement set using the task importvla. Standard
calibration was applied, with 3C286 serving as the flux calibrator and
3C287 as the phase calibrator. Imaging deconvolution was accomplished
with the task clean, where we used a multi-scale clean including
elements with a scale of 0, 5, 25, and 100 pixels. We used Briggs weighting
with a robust parameter of 0.5. The final image has a pixel scale of
0\farcs1 and an RMS of 0.28 mJy/beam at 1.41 GHz. The synthesized beam
is 1\farcs5$\times$1\farcs36, and the peak flux is 17 mJy/beam for the
radio core.

\begin{figure}
  \includegraphics[width=0.49\textwidth]{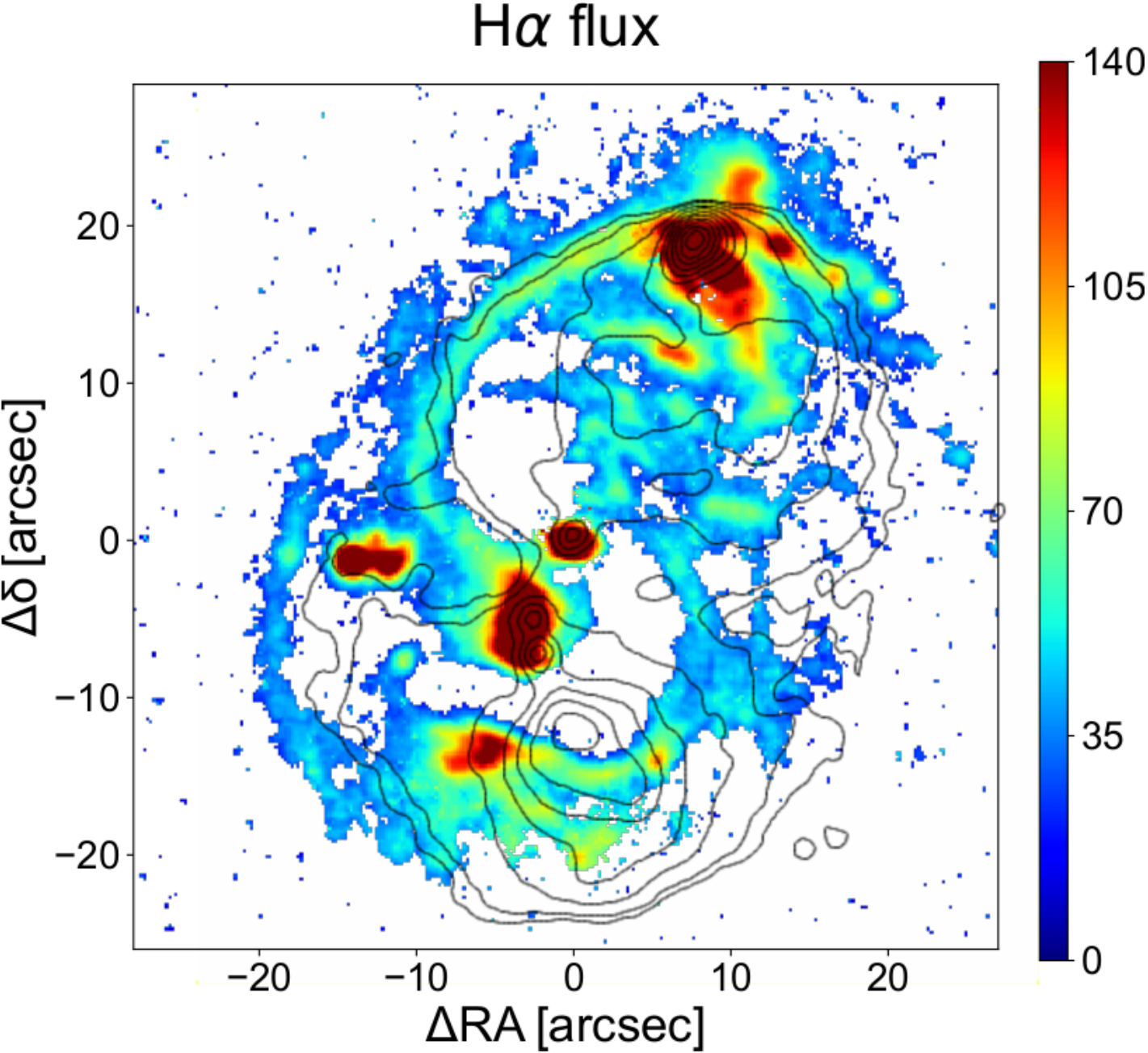}
  \caption{Distribution of the \Ha\ line emission where this is
      detected at a 2$\sigma$ level at least in single spaxels. The radio contours at 1.41 GHz are superposed. The lowest
      isocontour is at 0.9 mJy beam$^{-1}$. The contours then follow
      a geometric progression with a common ratio of 2.}
\label{composite}
\end{figure}

\begin{figure}
  \includegraphics[width=0.45\textwidth]{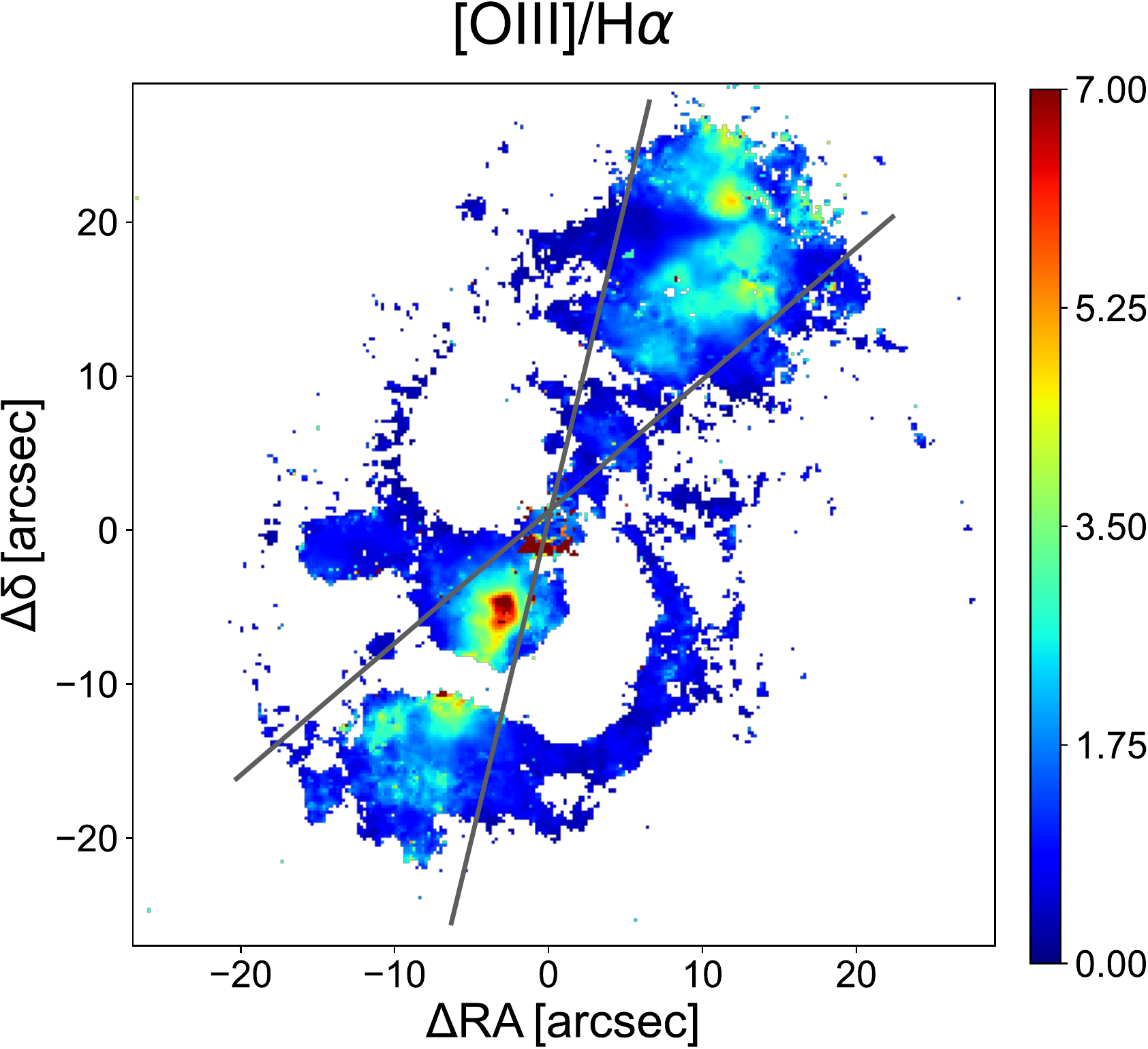}
  \caption{Ratio
    of the [O III] and \Ha\ lines where both lines are detected above
    a 2$\sigma$ level. The gray lines mark the boundary of the
    high-ionization bicone, i.e., the regions in which
    [O~III]/\Ha$>$2.5.}
\label{cono}
\end{figure}

\begin{figure*}
  \includegraphics[width=0.34\textwidth]{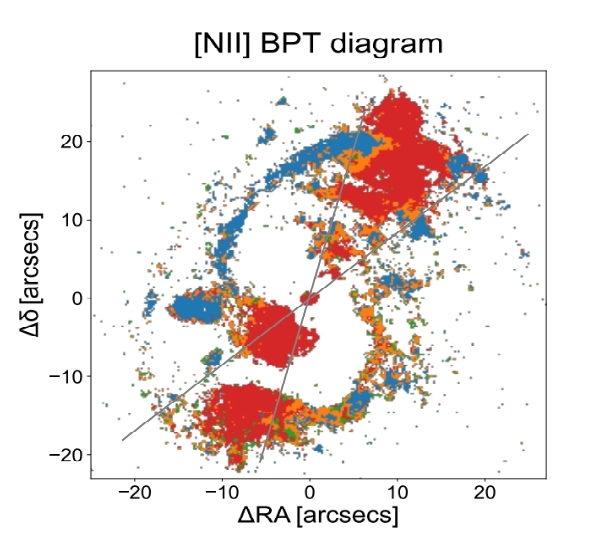}
    \includegraphics[width=0.337\textwidth]{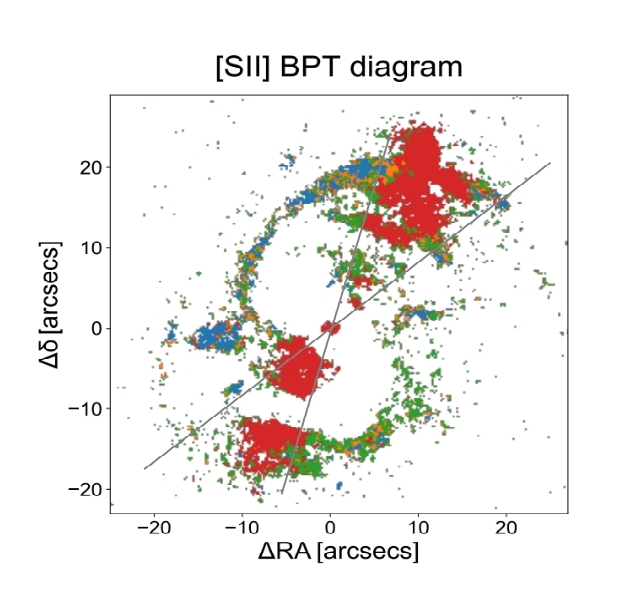}
  \includegraphics[width=0.355\textwidth]{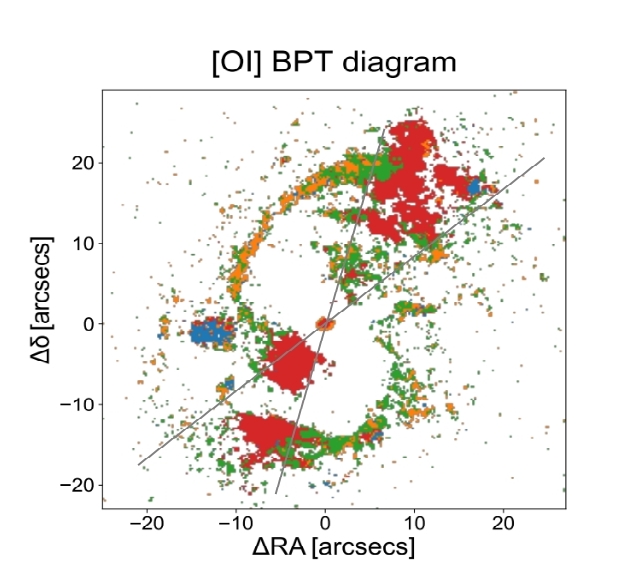}
   \includegraphics[width=0.34\textwidth]{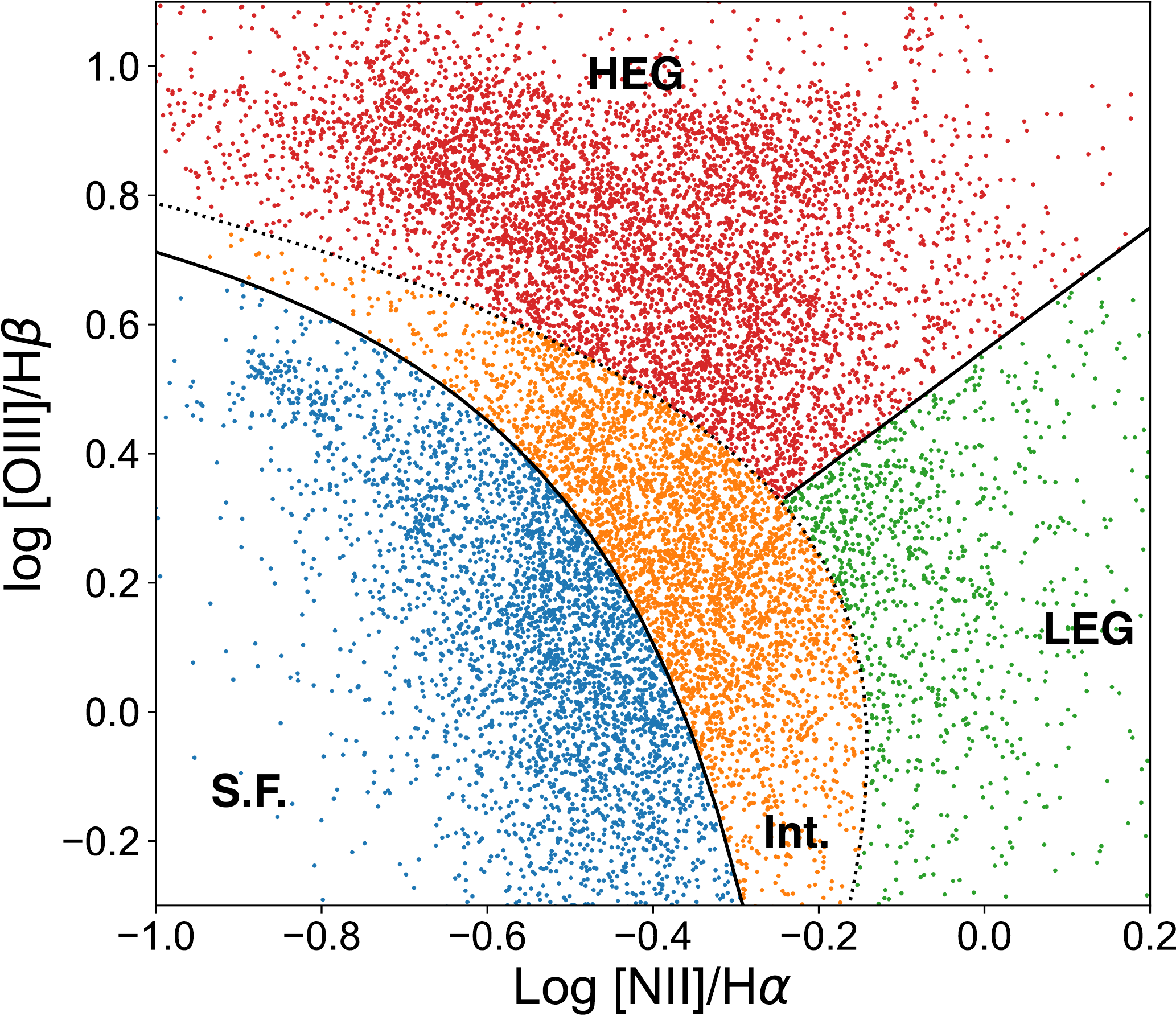}
    \includegraphics[width=0.33\textwidth]{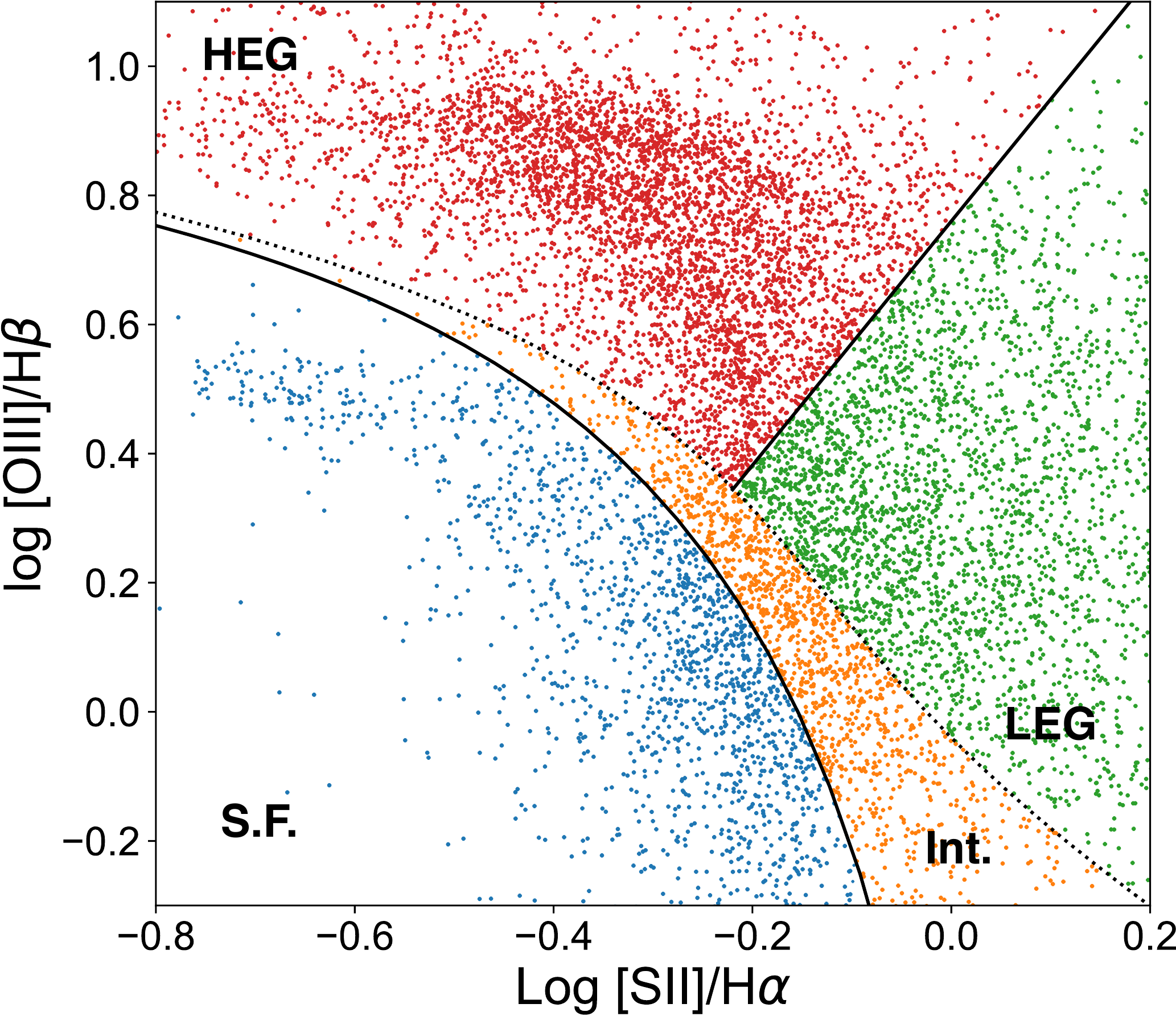}
  \includegraphics[width=0.32\textwidth]{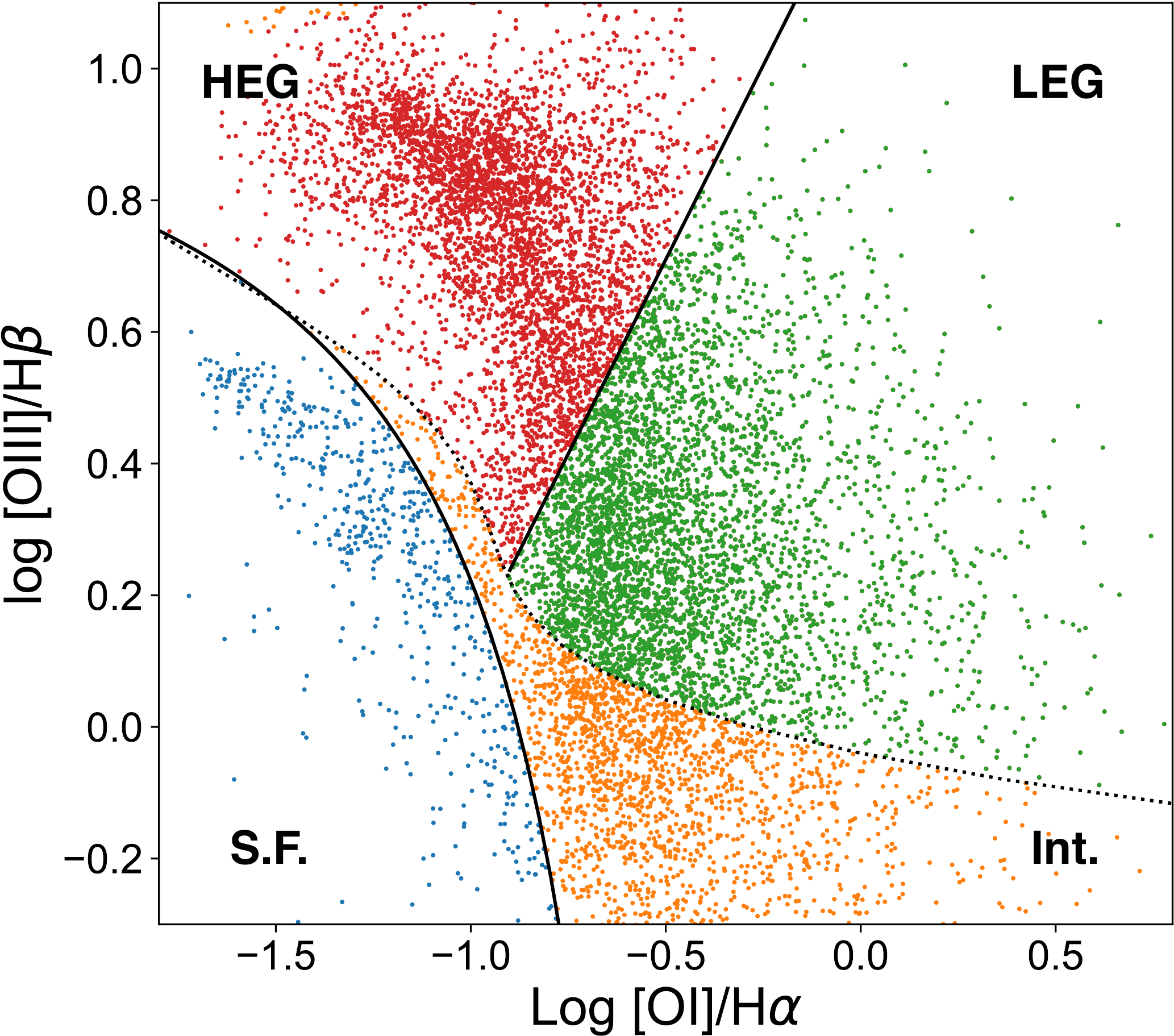}
  \caption{Images obtained by color-coding the emission line regions
    depending on the location of the representative points in the
    diagnostic diagrams. Red shows high-excitation regions, green
    shows low- excitation regions, blue shows star-forming regions,
    and orange represents intermediate objects using the regions from
    \citet{law20}. The images are defined where all four emission
    lines considered reach a 2$\sigma$ level in a single spaxel. The
    gray lines mark the boundary of the high-ionization bicone.}
\label{bpt}
\end{figure*}

\begin{figure}
  \includegraphics[width=0.45\textwidth]{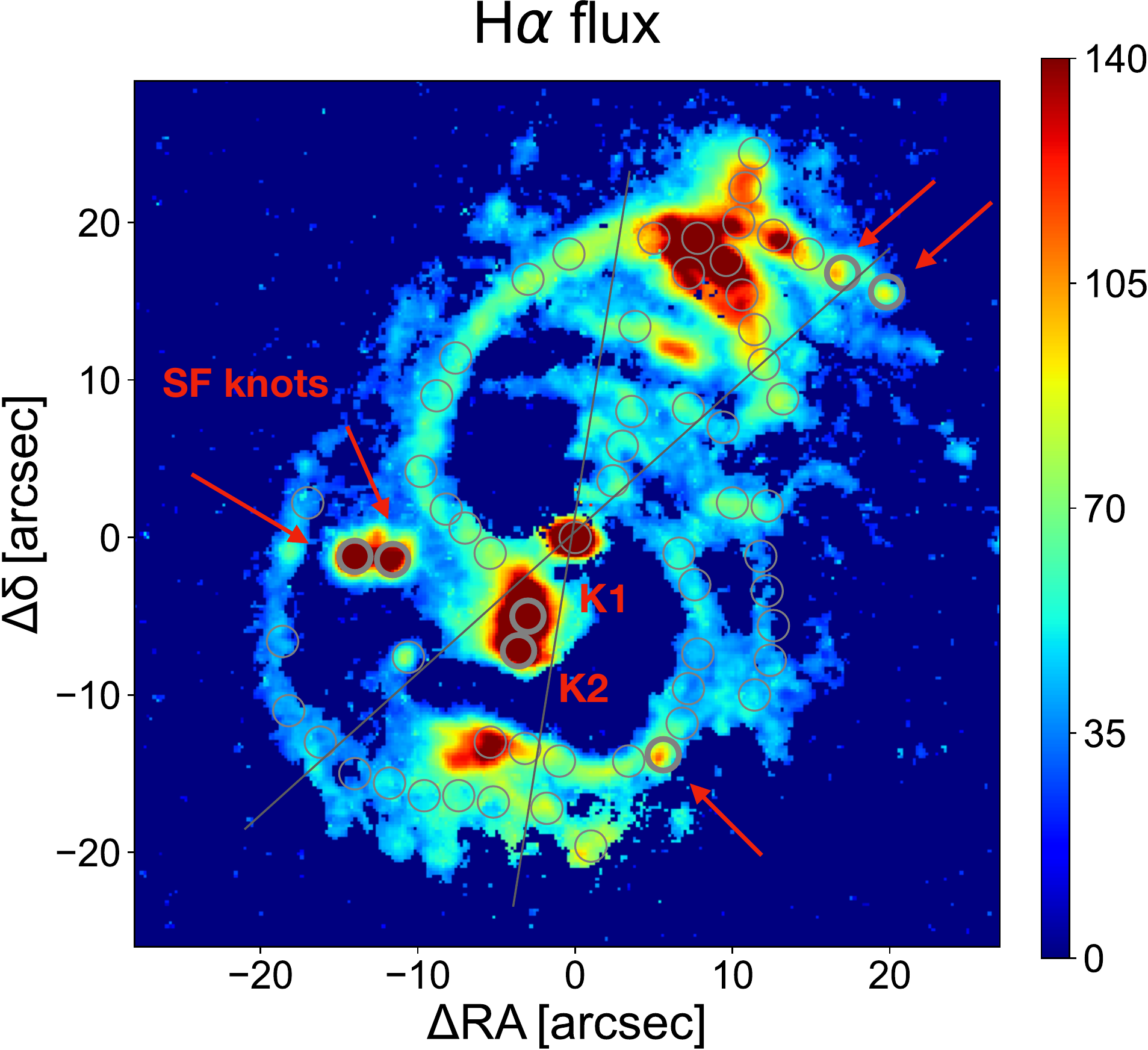}
  \caption{\Ha\ image of Coma A showing the regions in which this line is
    detected at a 2$\sigma$ level at least in single spaxels. We
    mark the locations from which we extracted the spectra (empty
    circles). We identify with red arrows the five compact knots of
    star formation and also mark the location of the radio knots K1
    and K2. The gray lines mark the boundary of the high
-ionization bicone. The color bar shown on the right side is in
    units of 10$^{-18}$ erg s$^{-1}$ cm$^{-2}$ arcsec$^{-2}$. }
\label{ha}
\end{figure}

\section{Results}
In Fig. \ref{composite} we show the distribution of the \Ha\ line
  emission. The radio contours are superposed. As already found
previously \citep{miley81,tadhunter00}, a nebula of ionized
gas with a size of $\sim 90$ kpc $\times 60$ kpc enshrouds the radio
emission. The brightest emission line regions are located along the
radio axis (at a position angle $\sim -25^\circ$), while a series of
narrow filaments follows the edges of the radio structure. 

In Fig. \ref{cono} we show an image of the ratio of the high-ionization line [O~III] with respect to the \Ha\ intensity\footnote{We
  preferred to produce the [O~III]/\Ha\ map instead of the reddening
-independent [O~III]/\Hb\ image because \Hb\ is detected at a
  sufficient S/N level over a much smaller area.} to study the
ionization structure within the nebula. This image shows the presence
of a biconical region characterized by a high [O~III]/\Ha\ ratio (as
high as $\sim 7$), with a sharp transition to significantly lower
ratios in the outward regions. By defining the bicone to include the
regions in which [O III]/\Ha$>$2.5, we obtain an axis at $\sim -30^\circ$
and an opening angle of $\sim 18^\circ$. The adopted threshold is
  somewhat arbitrary, but given the steep gradient in the
  [O~III]/\Ha\ ratio, values between 1.5 and 3.5 return similar values
  and suggest a rather small uncertainty in the cone size, $\sim
  5^\circ$.
  
\subsection{Gas ionization mechanism}

Spectroscopic diagnostic diagrams are commonly used to identify the
gas ionization mechanism
(e.g., \citealt{heckman80,baldwin81,veilleux87,kewley06,law20}). The
three diagrams are defined by four emission line ratios ([O
  III]/\Hb\ versus [N II]/\Ha, [S II]/\Ha, and [O I]/\Ha) that are
sensitive to the gas ionization properties. They allow us to identify
the ionization mechanism because gas ionized by star-forming regions
and by AGN fall into different regions of these diagrams.
In Fig. \ref{bpt} we show three images that are color-coded to identify
individual spaxels falling into the area of star-forming galaxies
(blue) and AGN, separating between high- and low-excitation galaxies
(HEGs in red and LEGs in green, respectively). Intermediate objects
are shown in orange, according to the regions defined by
\citet{law20}. These images are defined at the locations in which all
four emission lines we considered reach a 2$\sigma$ level in single
spaxels. They show a very complex ionization structure: a biconical
region, aligned with the radio axis, has a spectrum typical of HEGs,
and it is surrounded by filaments of lower ionization and regions with
spectra of star-forming regions.

To explore the ionization properties in more detail and explore
regions of low surface brightness that are not properly sampled by the single-spaxel analysis, we need to increase the S/N. We therefore extracted
spectra over circles with a radius of 1\farcs2 at the 76 locations
marked in Fig. \ref{ha}. We computed the emission line intensity
ratios for all regions. The lines S/N is sufficient (i.e., $>$3 for
all lines) to locate 63 of them (Fig. \ref{diag}). In
Fig. \ref{spectra} we show the spectra of four representative
regions. In the appendix, we list the position and line
ratios for these regions.

\begin{figure*}
  \includegraphics[width=0.99\textwidth]{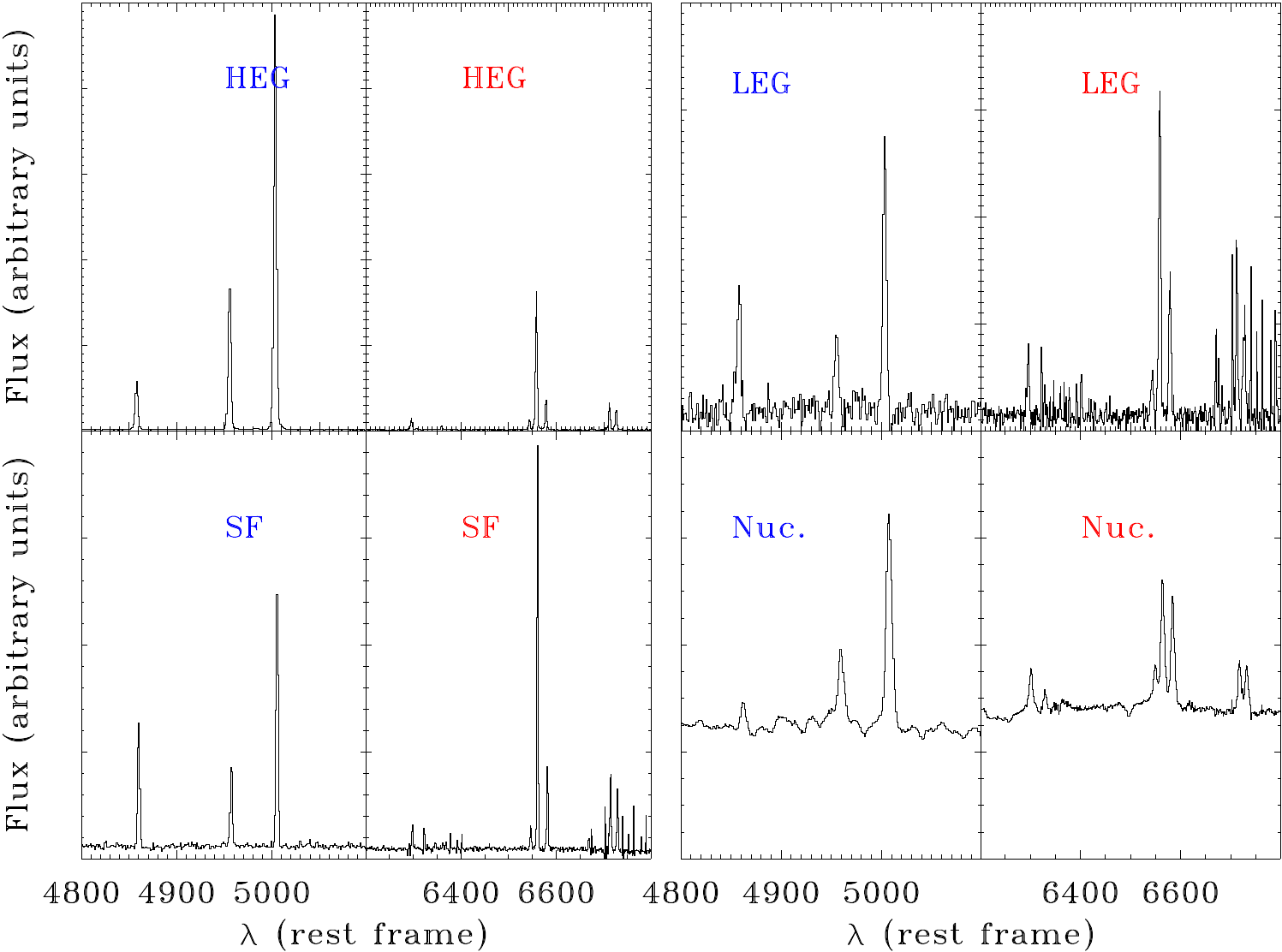}
  \caption{Blue and red spectra of four representative regions.}
\label{spectra}
\end{figure*}

\begin{figure*}
  \includegraphics[width=0.99\textwidth]{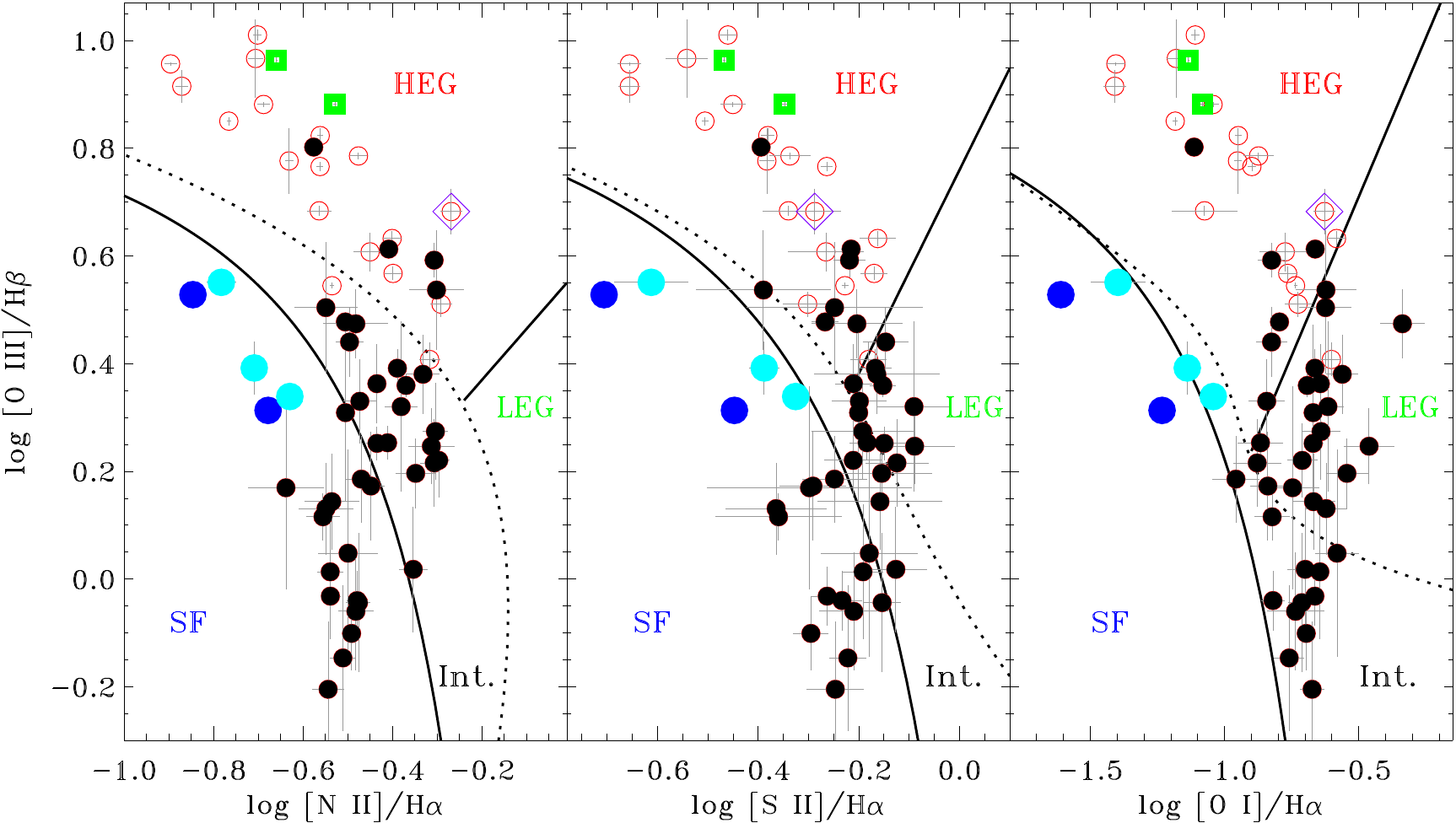}
  \caption{Location of the selected regions in the spectroscopic
    diagnostic diagrams. Filled black dots correspond to regions
    outside the high-ionization bicone, and the red empty dots are the
    regions inside the bicone (the nucleus is identified with an
    additional blue diamond). The five large dots correspond to the
    compact knots of star formation (the cyan dots are the three knots
    in the outer filaments). The green squares are associated with
    radio knots K1 and K2. The solid curves separate SF from
    intermediate galaxies (Int.), and the dotted curve separates
    intermediate galaxies from AGN. The black lines separate LEGs and
    HEGS \citep{law20}.}
\label{diag}
\end{figure*}

We identified five compact knots of line emission (marked in
Fig. \ref{diag} with blue and cyan circles) whose representative
points fall in the region of star-forming (SF) galaxies in all three
diagnostic diagrams. The two brightest knots are located east of the
nucleus, at a projected distance of $\sim 15$\arcsec\ ($\sim 25$
kpc). The other three are instead part of the large-scale filaments
visible in the NW and SW: the most distant of them is
$\sim$35\arcsec\ ($\sim 60$ kpc) NW of the host galaxy. The brightest
SF knot has a full width at half maximum that is consistent with the
seeing of the observations. The other four knots are superposed on
diffuse emission, and their sizes cannot be measured accurately.

The line ratios, in particular, the N2 ratios (-0.85$<$ log [N~II]/\Ha
$<$-0.6), are indicative of a subsolar gas metallicity, $Z \sim
0.6-0.8 Z_{\odot}$ \citep{pettini04}. The age of the stars can
instead be constrained from the equivalent width of the Balmer lines
\citep{leitherer99}. For the brightest SF knot, we measured an EW$_{\rm
  H\beta} = 80 \pm 3$ \AA,\ corresponding to an age of
$\sim3\times10^6$ years for an instantaneous burst of star
formation. As discussed in detail below, an instantaneous burst is the
most likely scenario in this source.

We estimated the effects of dust absorption by measuring the
\Ha/\Hb\ ratio in each region, adopting the \citet{cardelli89} law and
$R_V = 3.1$. The median value is E(B-V)$\sim$0.1 (with a range $0.05
\lesssim E(B-V) \lesssim 0.2$), which is larger than the Galactic value of
$\sim$0.01. This is indicative of significant internal reddening. There is no
clear structure in the gas reddening, nor a connection between the
reddening value and structures observed in the emission line flux maps.

We estimated the star formation rate (SFR) by measuring the
\Ha\ luminosity produced by the five SF knots. The resulting
\Ha\ luminosity, corrected for reddening, is $\sim 5.4\times10^{40}$
erg s$^{-1}$. The corresponding SFR is $\sim$0.8
M$_\odot$ y$^{-1}$. The dominant source of uncertainty for this
  estimate, a factor 3, is related to the adopted initial mass
  function of the stellar population \citep{pflamm07}.

\begin{figure*}
  \includegraphics[width=0.99\textwidth]{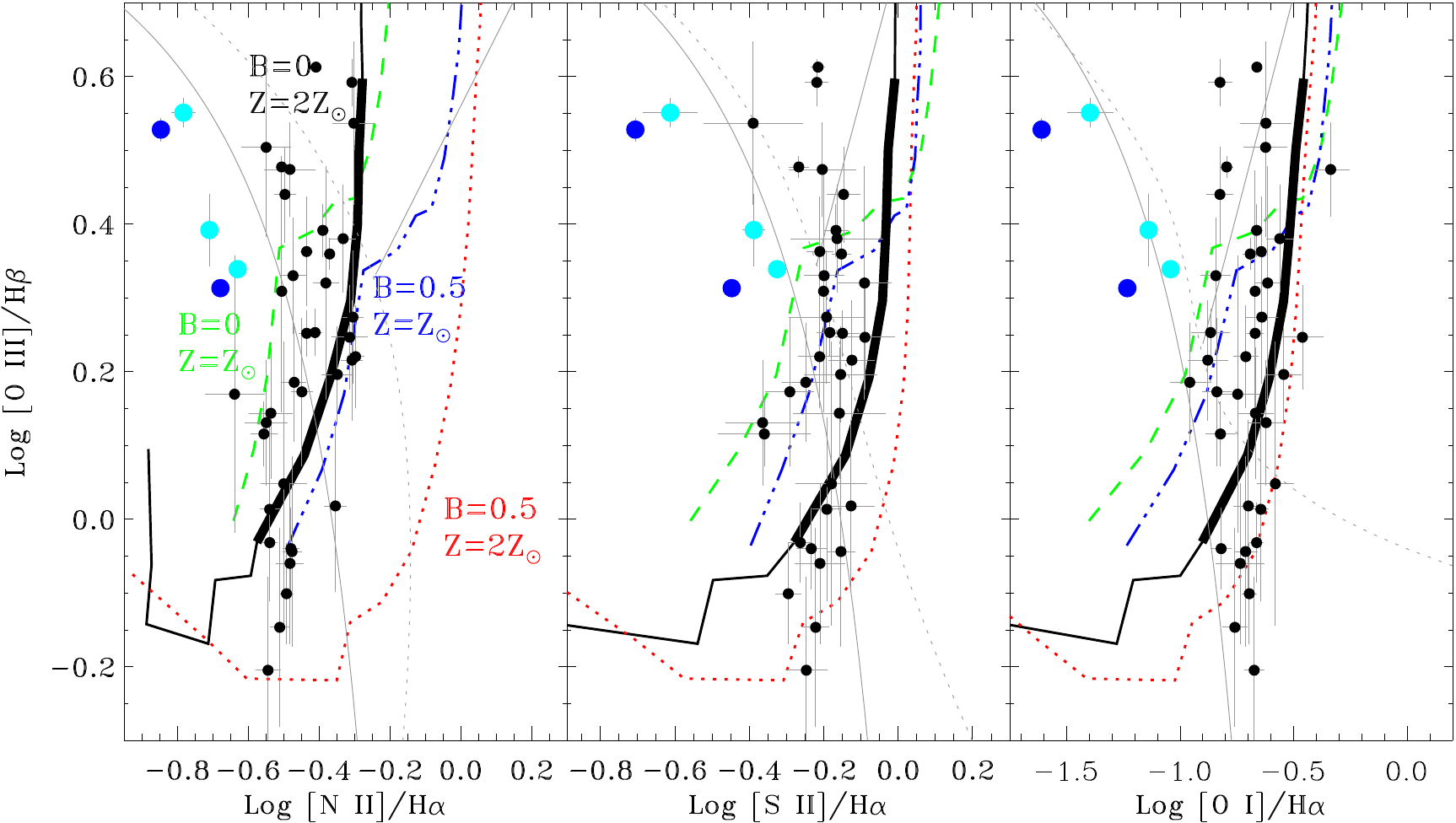}
  \caption{Comparison between the predictions from models of ionization
    from fast shocks \citep{allen08} and observed line ratios,
    limited to the regions outside the high-ionization bicone. The
    black curve shows the line ratios for a gas density of 1
    cm$^{-3}$, a metallicity $Z = 2 \times Z_{\sun}$, and a null
    magnetic field. The shock velocity ranges from 200 to 1000
    \kms\ in steps of 25 \kms, starting from the bottom left
    corner. The thick portion of the curve corresponds to the range
    $v_s = 350 - 500$ \kms. The dotted red lines are models with a
    magnetic field of 0.5 $\mu$G and $Z = 2 \times Z_{\sun}$. The
    green (dashed) and blue (dot-dashed) curves are the tracks obtained
    for a solar metallicity and B=0 $\mu$G and B=0.5 $\mu$G,
    respectively. The five large dots (coded as in Fig. 3) correspond
    to the compact knots of star formation (the cyan dots are the
    three knots in the outer filaments). }
\label{diagR}
\end{figure*}

The spectra of the majority of the selected locations within the high-ionization bicone
are characteristic of HEGs, the same behavior as in the
nucleus of Coma A. Regions with a higher [O III]/\Ha\ ratio
are aligned with the radio axis, but do not follow the curvature of the
radio source, in particular, in the south lobe. This suggests that in
this region, the main ionization mechanism is photoionization from the
nuclear radiation field, and it is not related to
the jets.

The regions outside the bicone show a complex behavior. In the first
and second diagnostic plots, they are mainly located in places in
which the line ratios correspond to those in SF galaxies
\citep{law20}. Some regions straddle the separation line between SF
galaxies and intermediate galaxies. This is suggestive of a mixed
contribution of star formation and AGN. However, in the third diagram,
they fully enter into the region of LEGs. This suggests that
photoionization, either by young stars or by an AGN, does not account
for the general location of the representative points of the
filamentary emission line regions in Coma A.

Several alternative gas ionization mechanisms other than
photoionization have been proposed: ionization due to hot cooling gas
\citep{voit94}, thermal conduction within the intracluster medium
\citep{mcdonald10}, reconnection diffusion \citep{fabian11}, and fast
ionizing shocks \citep{dopita95,allen08}. Only the shock models
produce an excitation state that is characterized by an [O III]/\Hb\ ratio as
high as that measured in the emission line filaments in Coma A.

We then focused on the MAPPINGS III library of shock models by
\citet{allen08}. We considered models of various metallicity, density,
and magnetic field. The model that best reproduces the observations,
that is, the model whose track follows the distribution of the measured
line ratios more closely, is shown in Fig. \ref{diagR}. It corresponds to a
twice solar metallicity, a gas density of 1 cm$^{-3}$, and a no
magnetic field. Overall, the predicted line ratios cover the range of
the observed values with shock velocities in the range 350 - 500 \kms.

The dependence of the line ratios on the various
parameters is strong. For example, for a magnetic field of 0.5$\mu$G, the
diagnostic tracks (the dotted red curves in Fig. \ref{diagR}) move
toward the bottom right corners of the diagrams and do not reproduce
the measured ratios. Similarly, the tracks obtained for a solar
metallicity (the green and blue curves in Fig. \ref{diagR}) overlap the observed values only marginally.

Finally, we estimated the density of the ionized gas by measuring the
[S~II]$\lambda$6716/[S~II]$\lambda$6731 ratio
\citep{osterbrock89}. Considering the errors (whose median value
  is 0.12), this ratio is always higher than 1.3. This is indicative of a
general low gas density ($n_e < 100$ cm$^{-3}$).

\begin{figure*}
  \includegraphics[width=0.51\textwidth]{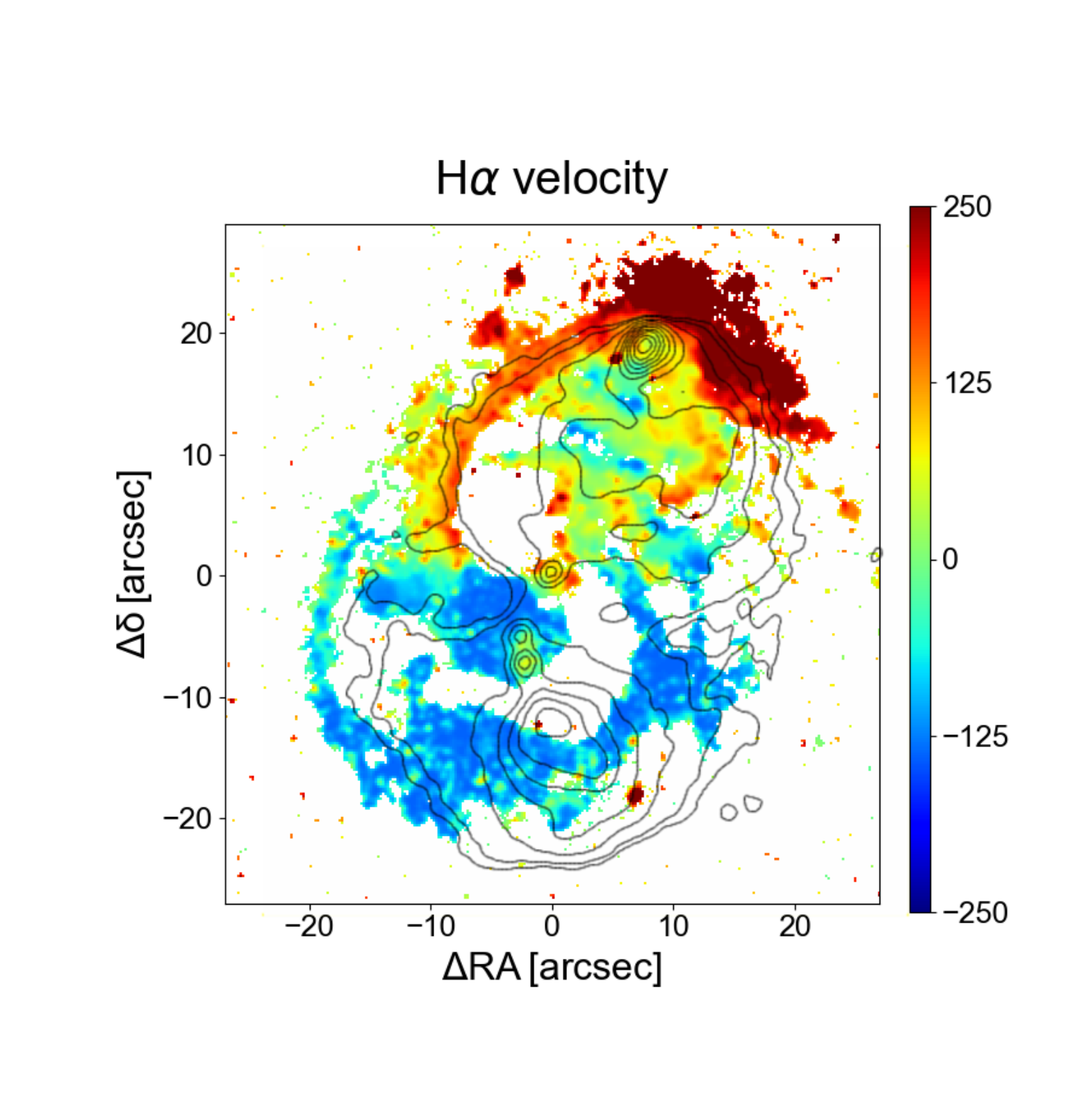}
  \includegraphics[width=0.52\textwidth]{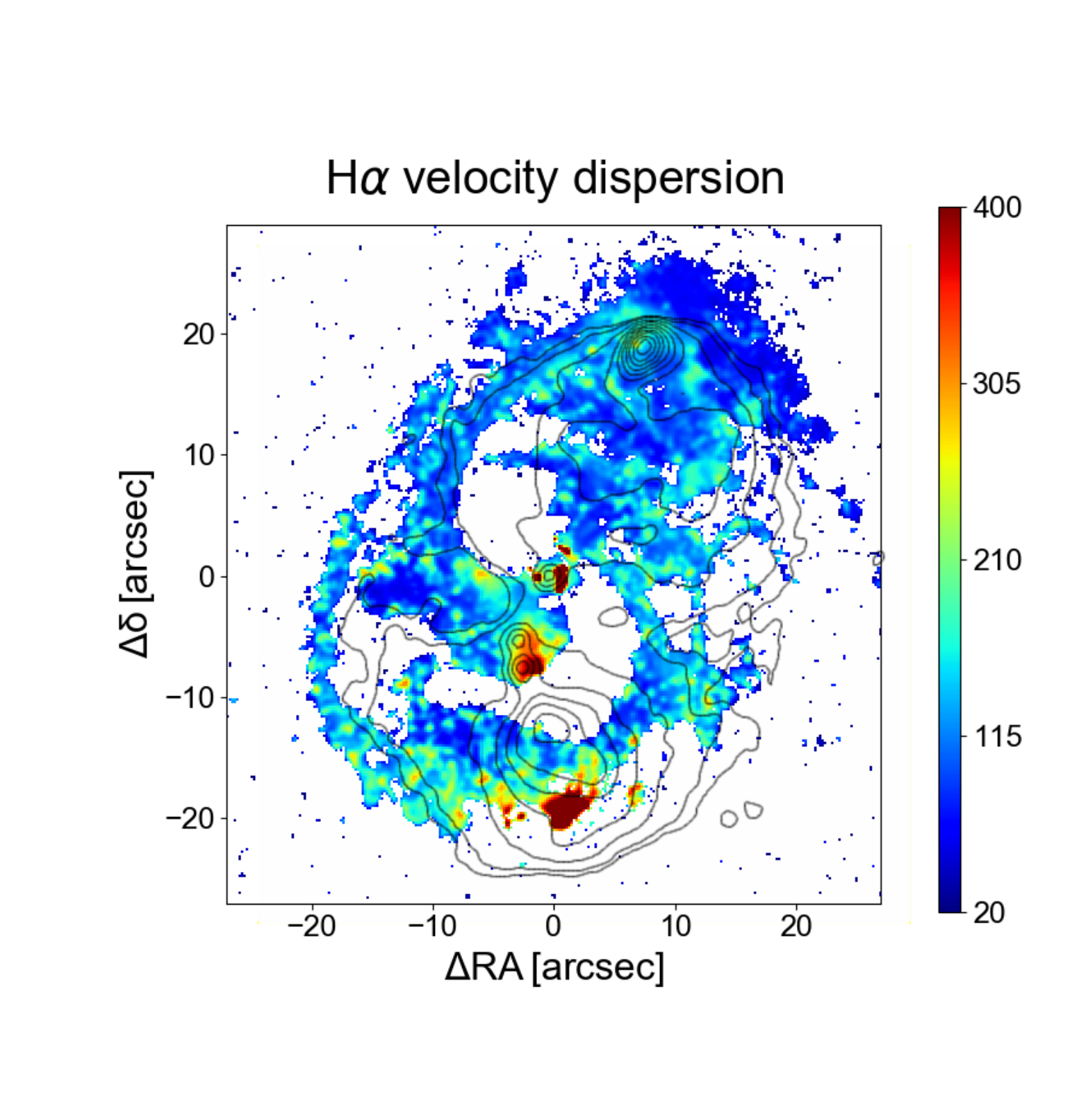}
  \caption{Velocity field (left) and velocity dispersion (not
    corrected for instrumental resolution, right), both in km
    s$^{-1}$, of the ionized gas derived from the
    \Ha\ emission line. We superposed the radio contours in both
    panels. Contours are drawn starting from 0.9 mJy beam$^{-1}$ with
    geometric progression with a common ratio of 2.}
\label{havel}
\end{figure*}

\subsection{Emission lines at the radio knots K1 and K2}

At a distance of $\sim 6\arcsec$ ($\sim10$ kpc) from the nucleus, the
southern jet bends by $\sim 40^\circ$. At this location lie two bright knots of radio emission (labeled K1 and K2 in
\citealt{vanbreugel85}). Emission lines associated with these knots
were first reported by \citet{miley81}, and nonthermal X-ray emission
is seen in the Chandra images \citep{worrall16}. The HST broadband
images \citep{martel99} show a narrow filament oriented along a NS
direction, and a similar morphology is seen in the [O~III] image
\citep{tilak05}. The emission line MUSE images
also present an elongated structure, but it is more extended, $\sim 5
\arcsec$, than that seen by HST ($\sim 2 \arcsec$ long).

The emission line ratios from the MUSE data for the two knots
(represented by the green squares in Fig. \ref{diag}) are similar to
those measured in the surrounding regions. The comparison with the
MAPPINGS III tracks does not return any shock model predicting the
observed ratios. These results suggest that despite the
in situ ionizing continuum seen in both the optical and X-ray images,
the dominant ionization mechanism in this region is due to the nuclear
radiation field.

\subsection{Gas kinematics}

Fig. \ref{havel} presents the velocity field of the ionized gas and
the distribution of line widths, not corrected for the
  instrumental resolution ($\sim 50$\kms\ at the \Ha\ wavelength, e.g.,
  \citealt{guerou17}). The velocities are referred to the redshift
measured ($z=0.08566$) in the nuclear spectrum from the stellar
absorption lines. The southern emission line lobe (associated with the
approaching radio jet) has a rather constant blueshifted velocity of
$\sim 100$ \kms\ with small fluctuations, the most notable being the
region immediately west of the radio knots K1 and K2, where a redshift of
$\sim 30$ \kms\ is observed. The northern lobe is instead generally
redshifted, but it shows a more complex velocity structure than the
southern lobe. The NE filament shows a redshift of $\sim 150$
\kms. Isolated blueshifted regions are also seen and are apparently connected
with the extension of filaments located in the southern lobe. The
highest velocities are reached beyond the edge of the northern radio
lobe, with velocities exceeding $\sim$ 300 \kms. Overall, the line
widths (not corrected for instrumental resolution) are between 100 and
150 \kms. The velocity dispersion reaches its maximum ($\gtrsim 400$
\kms) in two regions, at the southern tip of the nebula, close to
where the radio hot spot is located, and the other
$\sim$1\arcsec-2\arcsec\ just west of the K1 and K2 knot. It is
interesting that the region with the highest velocity dispersion is not
coincident with the location of the radio and line knots, but is offset from it
by a few kiloparsec. The lowest velocity widths (consistent with the MUSE
spectral resolution) are seen in the regions with the highest
velocities beyond the end of the northern lobe.

\section{Discussion}

From the analysis of the emission line ratios, we found evidence for
photoionization from young stars at five locations within the gas
nebula surrounding the radio emission in Coma A. As discussed by
\citet{mellema02}, a cloud overtaken by shocks breaks up into dense
fragments that are Jeans unstable, and might form stars. It then
appears that the shocks produced by the expansion of the radio lobes
are causing the formation of new stars. Star formation requires the
presence of dense clouds of cold gas. \citet{morganti02}
reported the detection of neutral hydrogen seen in absorption with a
total mass of at least $10^9$M$_\odot$ at distances as large as
$\sim$30 kpc from the center. Coma A is the only radio galaxy in which
H~I is seen in absorption at such large distances from the
nucleus. Furthermore, \citet{morganti02} suggested that the neutral
hydrogen and ionized gas, based on their similar kinematics, are part
of the same structure, possibly the remnant of a gas-rich merger.

Alternatively, the formation of molecular clouds can be due to thermal
instabilities that are causing the rapid cooling of gas that is outflowing
from the nucleus \citep{zubovas14}: a two-phase medium forms, with
cold dense molecular clumps mixed with hot tenuous gas, leading to
star formation. The situation in Coma A is somewhat different from
that envisaged by these authors because its outflow is probably
dominated by relativistic plasma, not by thermal gas. However, an
outflow of denser gas might be induced by a snow-plowing effect along
the edges of the radio source. 

For the remaining regions (outside the high-ionization bicone, marked
as black dots in Fig. \ref{diag} and \ref{diagR}) models of ionization
from fast shocks with velocities of 350-500 \kms\ are able to
reproduce the overall range of the measured emission line ratios. Despite
the reasonably good agreement between the model predictions and the
observations, the required metallicity value is higher than what we
found for the SF knots. furthermore, such a high metal content for gas
located at such large distances from the center of the host galaxy
appears to be rather contrived.

The detection of SF knots, in particular, those located along the
western filaments at the edges of the radio lobe, suggests that we
might be seeing the combined effect of shocks and young stars. 
  Gas ionized by the two processes is present within a given
  integration region, leading to the observed line ratios. A similar
mechanism of star formation within a turbulent flow has been suggested
by \citet{mcdonald12} to account for the emission line ratios observed
in the gaseous filaments of cool-core galaxy clusters.

The possibility of diffuse star formation within the filaments
suggests that the SFR value estimated above for the five star-forming
knots should be considered a lower limit. The total \Ha\ luminosity
outside the bicone is three times higher than that measured in the SF
knots alone. In addition, SF might occur also within the high-ionization bicone. However, the IRAS
satellite failed to detect Coma A: the upper limits in the 60 and 100
$\mu$m can be used to estimate a limit to the total far-IR luminosity of
L$_{\rm FIR}< 2.5 \times 10^{10}$ L$_\odot$ \citep{ocana10}, and
finally, using the relation by \citet{gao04}, a limit to the star
formation rate of about $<$5 M$_\odot$yr$^{-1}$.

{Previous studies of RGs show that these sources span a large
  range of SFR, beginning at the galaxy main sequence, but they are often
  found among the passive galaxies with very low SFRs
  \citep{dabhade20,bernhard21}. The stellar mass of Coma A can be
  estimated by using its K-band luminosity and the correlation from
  \citet{cappellari13}, resulting in $M_* = 3.3 \times 10^{11}
  M_\odot$. When combined with the estimate of the SFR, this value
  locates Coma A well within the galaxy main sequence
  \citep{saintonge17}. The key difference lies in the result that in
  Coma A, the star formation occurs well outside the host galaxy, at
  distances as large as 60 kpc.

The sharp edges of the radio lobes of 3C~277.3 indicate that they are
at a higher pressure than the external medium. The
expansion of an overpressured cocoon occurs at a supersonic speed
$v_{exp.} \sim c_s \times \sqrt{P_{c}/P_{em}}$, where $P_{c}$ and
$P_{em}$ are the pressure of the radio cocoon and of the external
medium, respectively \citep{begelman89}. The sound speed of the
external medium, $c_s$, for a temperature of 1 keV estimated from
X-ray images \citep{worrall16} is $\sim$ 500 \kms. We can set an upper
limit to the time $\tau_{exp.}$ in which the angular size of the radio
source grows by $\sim$ 0\farcs5, for instance (the resolution of the MUSE
data, corresponding to $dl = 800$ pc), $\tau_{exp.} = dl/v_{exp.} <
  1.6\times10^6 $ years. This timescale implies that we are
effectively seeing an instantaneous burst of star formation.

The expanding radio-emitting plasma also plays a role for the
kinematics of the ionized gas. As described above, the gas velocity
field is quite complex, and although it shows a general symmetry, it
is not compatible with gas in ordered rotation. At various locations
(e.g., immediately west of radio knots K1 and K2 and at the
southern radio hot spot), we found a connection between radio and
optical emission, with a strong increase in gas velocity
dispersion that bears witness to the effects of jet and cloud interactions. The
opposite behavior is seen in the region at the northern tip of the
ionized nebula, which is the only location in which gas is detected outside the
radio cocoon. Here, the gas has a small velocity gradient and produces
line emission with a very small width. This indicates that it is
unperturbed. However, if the expansion of the lobes were the dominant
driver of the gas kinematics and this occurred within a gas structure
encompassing the whole radio structure, we would expect to see a quite
different velocity field: the two lobes should each induce the
expansion of a gas bubble, with a blueshifted and a redshifted
component on either side. The observed kinematics can be reconciled
with a dominance of the radio outflows if the gas has a flattened disk
structure and if the jets are oriented so that they graze the gas
disk: in this case, only one side of the lobe would interact with the
denser gas regions, producing the observed asymmetry.  Alternatively,
the gas dynamics is dominated by gravity and the radio outflow only
produces localized disturbances, or the complex velocity field is due
to the external (merger) origin of the gas that has not yet reached a
dynamical equilibrium.

The brightest regions of line emission in Coma A are located within a
biconical region in which we also find the highest
[O~III]/\Ha\ ratios. This morphology is reminiscent of what is often
observed in type 2 Seyfert galaxies (e.g., \citealt{tadhunter89}), and
it is indicative of  circumnuclear selective
obscuration, as postulated by the unified model (UM) of AGN
\citep{antonucci93}. The lack of broad lines in Coma A
\citep{buttiglione10} and the high column density derived from the
nuclear X-ray spectrum \citep{worrall16,macconi20} are in line with
the idea that our view toward its nucleus intercepts regions of high
absorption. However, the angular size of the bicone is smaller ($\sim
18^\circ\pm5^\circ$) than predicted by the UM estimate: based on the
relative number of 3CR RGs with $z<0.3$ with and without broad lines
in their optical spectra. \citet{baldi13} estimated that this fraction
corresponds to an average cone aperture of $\sim 50\pm5^\circ$. This
discrepancy might be due to the combination of an intrinsic anisotropy
of the nuclear radiation field and additional
ionization mechanisms. In this case, the relative contribution of the
AGN light is reduced at larger angles from the disk axis. The nuclear
anisotropy might be due to the geometrical thickness of the accretion
disk (e.g., \citealt{sikora81}), but also to a contribution to the
radiation field from relativistically beamed emission from the jet
base. While in blazars, this collimated beam points at a small angle
from the line of sight, in radio galaxies such as Coma A, it would
illuminate a biconical region with an opening angle $\theta \sim
\Gamma^{-1}$, where $\Gamma$ is the jet Lorentz factor.

\section{Summary and conclusions}

We presented the results of VLT/MUSE observations of the nearby radio
galaxy Coma A that show a large nebula ($\sim 90$ kpc $\times 60$ kpc)
of ionized gas cospatial with the radio emission. We estimated the
emission line ratios in several regions throughout the nebula to explore
the gas ionization mechanism. Five compact knots have line ratios that
are indicative of ionization from young stars with an estimated age of
$\sim3\times10^6$ years and a subsolar gas metallicity, $Z \sim
0.6-0.8 Z_{\odot}$.

Three star-forming regions (detected as far as 60 kpc from the host)
are part of the gas filaments surrounding the western edges of the radio
lobes. The most likely origin of these filaments is the compression of
the ambient gas produced by the expanding radio source that increases
its density, boosting the line emission. This same compression causes
the collapse of the gas clouds, leading to the formation of new stars.
The two brightest SF knots are instead located at a projected distance
of 25 kpc east of the center of the galaxy, within a channel of lower
radio surface brightness between the two lobes. In this region, the
compression might be due to the plasma flowing back from the radio
spots. There is not necessarily a causal connection between
the radio ejecta and the star formation event at this location: star-forming knots are commonly associated with gas-rich merger events, the
most likely explanation for the large amount of gas in
Coma A.

A reservoir of dense gas is needed to maintain star formation. Cold
gas is indeed detected in Coma A from its HI absorption against the
radio continuum extending to a very large radius, at least $\sim 30$
kpc.  In this context, it would be of great interest to derive the
distribution and kinematics of the molecular gas and its connection
with the radio-emitting and ionized gas. The clumps of star formation
found at $\sim$40-60 kpc from the center of the galaxy suggest that
the molecular gas should extend to these large distances. Its
distribution would enable us to test whether it survives within the
radio lobes or is destroyed by shock heating, ultimately quenching
future star formation.

In addition to the star-forming regions, the gas nebula around Coma A shows a
well-defined ionization structure. Within a narrow bicone, with an
opening angle of $\sim$18$^\circ$, the gas is in a high-ionization state. The relatively small angle of the bicone (with
respect to the indications of the AGN unification scheme) suggests
that this is not solely due to circumnuclear obscuration, but that an
intrinsic anisotropy of the nuclear radiation field is present. This
might be due to effects of radiative transport within a thick
accretion disk, but there is also the possibility that we are seeing the
result of relativistic beaming, that is, that a blazar nucleus, oriented
at a large angle from our line of sight, is present in this radio
source.

Outside the high-ionization bicone, the gas is in a much lower
ionization state. The location in the diagnostic diagrams does not
follow the distribution of photoionized gas from either an active
nucleus or young stars, suggesting an additional
ionization mechanism. Shocks with a velocity of 350-500 \kms\ produce
line ratios similar to the observed ratios, but with preferred values of
null magnetic field and a twice solar gas metallicity. These values
contradict the measurements based on the Faraday rotation ($\sim 1
\mu$G) and with the metallicity estimated on the SF knots. This
suggests a contribution from diffuse star formation produced within
the ionized gas filaments. UV imaging represents the best tool to
separate the contribution of SF and shocks to the gas ionization
budget because as shown by \citet{mcdonald11}, the ratio between line
and UV continuum emission is radically different in these two
scenarios. UV images would also enable us to map the distribution and
estimate the age of young stars, and to follow the star formation
history in this galaxy. Given the high advancing speed of the radio
lobes, they might leave a resolved trail of star-forming regions of
increasing age when they move toward the nucleus.

In low-redshift RGs with evidence of star formation triggered by
the jets discussed in the Introduction, young stars are formed in a
few locations along the path or at the termination of a radio
jet. In these objects, the SFR is generally lower than our
  conservative estimate for Coma A ($\sim$0.8 M$_\odot$ y$^{-1}$):
  1$\times10^{-3}-5\times10^{-3}$ M$_\odot$ y$^{-1}$ in Centaurus A
  \citep{salome16}, and 0.5, and 0.03 for the Minkowski object, and
NGC~5643, respectively. However, the main distinguishing feature in
this source is that the line ratios suggest that star formation,
clearly detected in a few individual knots, occurs throughout the whole
gaseous structure that enshrouds the whole radio source at distances as
large as 60 kpc. In this sense, Coma A might be a unique source in the
local Universe in which the expanding radio source appears to be
triggering a global event of star formation. This represents an excellent
laboratory for exploring the mechanism of positive feedback in detail.

Based on observations collected at the European Southern Observatory
under ESO program 0102.B-0048(A).

S.B. and C.OD. are grateful to the Natural Sciences and
Engineering Research Council (NSERC) of Canada.

R.G. acknowledges support from the agreement ASI-INAF n. 2017-14-H.O
G.V. acknowledges support from ANID program FONDECYT Postdoctorado 3200802.

\clearpage
\newpage
\begin{appendix}

  \section{Emission line ratio measurements}
  \begin{table*}
\caption{Emission line ratio measurements}
\label{appA}  
\begin{tabular}{r r r r r r l}
\hline
R.A.& Dec. &[O~III]/\Hb & [N~II]/\Ha & [S~II]/\Ha & [O~I]/\Ha & Region\\ 
\hline
  19.6 &  15.4 &  0.55$\pm$ 0.02 & -0.78 $\pm$0.03  & -0.61$\pm$ 0.07  & -1.40$\pm$ 0.10 &  SF\\ 
  16.8 &  16.6 &  0.39$\pm$ 0.05 & -0.71 $\pm$0.02  & -0.39$\pm$ 0.03  & -1.14$\pm$ 0.04 &  SF\\ 
  14.6 &  17.8 &  0.88$\pm$ 0.01 & -0.69 $\pm$0.01  & -0.45$\pm$ 0.02  & -1.04$\pm$ 0.02 &  \\ 
  13.0 &   8.6 &  0.01$\pm$ 0.12 & -0.54 $\pm$0.03  & -0.19$\pm$ 0.04  & -0.64$\pm$ 0.04 &  \\ 
  12.4 &  19.0 &  0.96$\pm$ 0.01 & -0.90 $\pm$0.01  & -0.66$\pm$ 0.02  & -1.40$\pm$ 0.04 &  \\ 
  12.4 &  -5.8 &  0.20$\pm$ 0.07 & -0.35 $\pm$0.05  & -0.15$\pm$ 0.10  & -0.54$\pm$ 0.08 &  \\ 
  12.2 &  -8.0 & -0.06$\pm$ 0.11 & -0.48 $\pm$0.04  & -0.21$\pm$ 0.07  & -0.74$\pm$ 0.11 &  \\ 
  12.0 &   1.8 &  0.19$\pm$ 0.08 & -0.47 $\pm$0.03  & -0.25$\pm$ 0.06  & -0.96$\pm$ 0.09 &  \\ 
  12.0 &  -3.6 &  0.25$\pm$ 0.07 & -0.31 $\pm$0.05  & -0.09$\pm$ 0.08  & -0.46$\pm$ 0.09 &  \\ 
  11.8 &  10.8 &  0.48$\pm$ 0.01 & -0.51 $\pm$0.02  & -0.27$\pm$ 0.03  & -0.80$\pm$ 0.03 &  \\ 
  11.6 &  -1.4 &  0.54$\pm$ 0.11 & -0.30 $\pm$0.06  & -0.39$\pm$ 0.13  & -0.62$\pm$ 0.11 &  \\ 
  11.2 &  24.2 &  0.97$\pm$ 0.07 & -0.71 $\pm$0.02  & -0.54$\pm$ 0.04  & -1.18$\pm$ 0.04 &  \\ 
  11.2 &  13.0 &  0.80$\pm$ 0.02 & -0.58 $\pm$0.01  & -0.39$\pm$ 0.02  & -1.11$\pm$ 0.03 &  \\ 
  11.2 & -10.2 &  0.02$\pm$ 0.12 & -0.35 $\pm$0.03  & -0.13$\pm$ 0.06  & -0.70$\pm$ 0.07 &  \\ 
  10.6 &  22.0 &  1.01$\pm$ 0.01 & -0.70 $\pm$0.01  & -0.46$\pm$ 0.02  & -1.11$\pm$ 0.02 &  \\ 
  10.4 &  15.2 &  0.92$\pm$ 0.03 & -0.87 $\pm$0.01  & -0.66$\pm$ 0.02  & -1.41$\pm$ 0.04 &  \\ 
  10.2 &  19.8 &  0.77$\pm$ 0.01 & -0.56 $\pm$0.01  & -0.26$\pm$ 0.01  & -0.90$\pm$ 0.01 &  \\ 
   9.8 &   2.0 & -0.04$\pm$ 0.06 & -0.48 $\pm$0.02  & -0.23$\pm$ 0.04  & -0.82$\pm$ 0.05 &  \\ 
   9.4 &  17.4 &  0.85$\pm$ 0.01 & -0.77 $\pm$0.01  & -0.51$\pm$ 0.01  & -1.18$\pm$ 0.01 &  \\ 
   9.2 &   6.8 &  0.13$\pm$ 0.09 & -0.55 $\pm$0.06  & -0.36$\pm$ 0.10  & -0.62$\pm$ 0.08 &  \\ 
   7.6 &  18.8 &  0.54$\pm$ 0.01 & -0.54 $\pm$0.01  & -0.23$\pm$ 0.01  & -0.74$\pm$ 0.01 &  \\ 
   7.6 &  -7.6 &  0.38$\pm$ 0.07 & -0.33 $\pm$0.04  & -0.16$\pm$ 0.13  & -0.56$\pm$ 0.06 &  \\ 
   7.4 &  -3.2 &  0.27$\pm$ 0.09 & -0.30 $\pm$0.03  & -0.19$\pm$ 0.10  & -0.64$\pm$ 0.07 &  \\ 
   7.0 &  16.6 &  0.78$\pm$ 0.06 & -0.63 $\pm$0.01  & -0.38$\pm$ 0.01  & -0.95$\pm$ 0.01 &  \\ 
   7.0 &   8.0 &  0.17$\pm$ 0.10 & -0.45 $\pm$0.03  & -0.29$\pm$ 0.07  & -0.84$\pm$ 0.07 &  \\ 
   7.0 &  -9.8 &  0.32$\pm$ 0.16 & -0.38 $\pm$0.04  & -0.09$\pm$ 0.07  & -0.61$\pm$ 0.06 &  \\ 
   6.6 & -12.0 &  0.44$\pm$ 0.07 & -0.50 $\pm$0.03  & -0.15$\pm$ 0.05  & -0.82$\pm$ 0.06 &  \\ 
   6.4 &  -1.2 &  0.22$\pm$ 0.08 & -0.31 $\pm$0.03  & -0.12$\pm$ 0.06  & -0.88$\pm$ 0.09 &  \\ 
   5.4 & -14.0 &  0.34$\pm$ 0.04 & -0.63 $\pm$0.02  & -0.33$\pm$ 0.04  & -1.04$\pm$ 0.04 &  SF\\ 
   4.8 &  18.8 &  0.31$\pm$ 0.16 & -0.50 $\pm$0.01  & -0.20$\pm$ 0.01  & -0.67$\pm$ 0.01 &  \\ 
   3.6 &  13.2 &  0.57$\pm$ 0.01 & -0.40 $\pm$0.02  & -0.17$\pm$ 0.03  & -0.76$\pm$ 0.04 &  \\ 
   3.4 &   7.8 &  0.41$\pm$ 0.03 & -0.32 $\pm$0.02  & -0.18$\pm$ 0.04  & -0.60$\pm$ 0.04 &  \\ 
   3.2 & -14.4 &  0.25$\pm$ 0.03 & -0.43 $\pm$0.02  & -0.15$\pm$ 0.05  & -0.67$\pm$ 0.04 &  \\ 
   2.8 &   5.6 &  0.51$\pm$ 0.02 & -0.29 $\pm$0.02  & -0.30$\pm$ 0.05  & -0.73$\pm$ 0.05 &  \\ 
   1.0 & -15.0 &  0.36$\pm$ 0.07 & -0.43 $\pm$0.02  & -0.21$\pm$ 0.04  & -0.64$\pm$ 0.03 &  \\ 
   0.8 & -19.8 &  0.25$\pm$ 0.03 & -0.41 $\pm$0.02  & -0.18$\pm$ 0.03  & -0.87$\pm$ 0.08 &  \\ 
   0.0 &   0.0 &  0.68$\pm$ 0.04 & -0.27 $\pm$0.02  & -0.29$\pm$ 0.05  & -0.63$\pm$ 0.04 &  Nuc.\\ 
  -0.6 &  17.8 & -0.03$\pm$ 0.05 & -0.54 $\pm$0.02  & -0.26$\pm$ 0.03  & -0.66$\pm$ 0.03 &  \\ 
  -1.2 & -14.4 &  0.39$\pm$ 0.04 & -0.39 $\pm$0.02  & -0.17$\pm$ 0.03  & -0.66$\pm$ 0.03 &  \\ 
  -1.6 &  -2.2 &  1.29$\pm$ 0.05 & -0.23 $\pm$0.02  & -0.33$\pm$ 0.06  & -0.75$\pm$ 0.05 &  \\ 
  -2.0 & -17.4 &  0.36$\pm$ 0.02 & -0.37 $\pm$0.01  & -0.15$\pm$ 0.03  & -0.69$\pm$ 0.03 &  \\ 
  -3.2 &  16.2 & -0.10$\pm$ 0.07 & -0.49 $\pm$0.02  & -0.30$\pm$ 0.04  & -0.70$\pm$ 0.03 &  \\ 
  -3.2 &  -5.2 &  0.96$\pm$ 0.01 & -0.66 $\pm$0.01  & -0.47$\pm$ 0.01  & -1.13$\pm$ 0.01 &  K1\\ 
  -3.4 & -13.6 &  0.61$\pm$ 0.01 & -0.41 $\pm$0.01  & -0.22$\pm$ 0.02  & -0.66$\pm$ 0.02 &  \\ 
  -3.8 &  -7.4 &  0.88$\pm$ 0.01 & -0.53 $\pm$0.01  & -0.35$\pm$ 0.01  & -1.08$\pm$ 0.01 &  K2\\ 
  -5.4 & -17.0 &  0.63$\pm$ 0.01 & -0.40 $\pm$0.02  & -0.16$\pm$ 0.04  & -0.58$\pm$ 0.04 &  \\ 
  -5.6 &  -1.2 &  0.59$\pm$ 0.03 & -0.31 $\pm$0.02  & -0.22$\pm$ 0.03  & -0.82$\pm$ 0.05 &  \\ 
  -5.6 & -13.2 &  0.82$\pm$ 0.01 & -0.56 $\pm$0.01  & -0.38$\pm$ 0.01  & -0.95$\pm$ 0.01 &  \\ 
  -7.2 &   0.4 &  0.22$\pm$ 0.07 & -0.30 $\pm$0.02  & -0.21$\pm$ 0.06  & -0.71$\pm$ 0.06 &  \\ 
  -7.6 & -16.6 &  0.79$\pm$ 0.01 & -0.48 $\pm$0.02  & -0.34$\pm$ 0.04  & -0.87$\pm$ 0.06 &  \\ 
  -7.8 &  11.2 & -0.20$\pm$ 0.13 & -0.54 $\pm$0.04  & -0.25$\pm$ 0.06  & -0.67$\pm$ 0.04 &  \\ 
  -9.0 &   8.8 & -0.15$\pm$ 0.14 & -0.51 $\pm$0.03  & -0.22$\pm$ 0.04  & -0.76$\pm$ 0.06 &  \\ 
  -9.8 & -16.6 &  0.68$\pm$ 0.01 & -0.56 $\pm$0.03  & -0.34$\pm$ 0.05  & -1.07$\pm$ 0.12 &  \\ 
 -10.0 &   4.0 & -0.04$\pm$ 0.13 & -0.48 $\pm$0.03  & -0.15$\pm$ 0.04  & -0.71$\pm$ 0.05 &  \\ 
 -10.8 &  -7.8 &  0.12$\pm$ 0.04 & -0.56 $\pm$0.04  & -0.36$\pm$ 0.13  & -0.82$\pm$ 0.07 &  \\ 
 -11.8 &  -1.6 &  0.31$\pm$ 0.01 & -0.68 $\pm$0.01  & -0.45$\pm$ 0.02  & -1.23$\pm$ 0.02 &  SF\\ 
 -12.0 & -15.8 &  0.61$\pm$ 0.04 & -0.45 $\pm$0.04  & -0.27$\pm$ 0.07  & -0.77$\pm$ 0.10 &  \\ 
 -14.2 &  -1.4 &  0.53$\pm$ 0.02 & -0.85 $\pm$0.01  & -0.71$\pm$ 0.01  & -1.61$\pm$ 0.02 &  SF\\ 
 -14.2 & -15.2 &  0.47$\pm$ 0.06 & -0.48 $\pm$0.07  & -0.20$\pm$ 0.09  & -0.34$\pm$ 0.08 &  \\ 
 -16.4 & -13.2 &  0.14$\pm$ 0.09 & -0.54 $\pm$0.06  & -0.16$\pm$ 0.12  & -0.67$\pm$ 0.08 &  \\ 
 -17.2 &   2.0 &  0.17$\pm$ 0.19 & -0.64 $\pm$0.08  & -0.30$\pm$ 0.20  & -0.75$\pm$ 0.10 &  \\ 
 -18.4 & -11.2 &  0.50$\pm$ 0.12 & -0.55 $\pm$0.07  & -0.25$\pm$ 0.18  & -0.62$\pm$ 0.10 &  \\ 
 -18.8 &  -6.8 &  0.05$\pm$ 0.19 & -0.50 $\pm$0.07  & -0.18$\pm$ 0.10  & -0.58$\pm$ 0.08 &  \\ 
\hline
\end{tabular}

\medskip
Column description: (1 and 2) offset, in arcseconds, from the nucleus, (3-6) diagnostic emission line ratios, (7) region. 
\end{table*}

\end{appendix}

\end{document}